\documentclass[preprint,tightenlines,eqsecnum,aps,epsfig]{revtex4}
\usepackage{epsfig}

\def\beq{\begin{equation}}   \def\eeq{
\end{equation}}

\begin{document}
\title{ On the behavior of single scale hard small
$x$ processes in  QCD near the black disc limit }
\author{ B. Blok\email{E-mail: blok@physics.technion.ac.il} }
\affiliation{Department of Physics, Technion---Israel Institute of
Technology, 32000 Haifa, Israel}
\author{ L. Frankfurt\email{E-mail: frankfur@lev.tau.ac.il} }
\affiliation{School of Physics and Astronomy, Raymond and Beverly Sackler
Faculty of Exact Sciences,
Tel Aviv University, 69978 Tel Aviv,
Israel}

\thispagestyle{empty}

\begin{abstract}

We argue that at sufficiently small Bjorken $x$ where pQCD
amplitudes rapidly increase with energy and violate probability
conservation the shadowing effects in the single-scale small $x$
hard QCD processes can be described by an effective  quantum field
theory of interacting quasiparticles-- perturbative QCD ladders.
We find, within the WKB approximation, that the smallness of the
QCD coupling constant ensures the hierarchy among
many-quasiparticle interactions evaluated within the physical
vacuum and in particular, the dominance  in the Lagrangian of the
triple quasiparticle interaction. It is explained that the
effective field theory considered near the perturbative QCD vacuum
contains a  tachyon relevant for the divergency of the
perturbative QCD series at sufficiently small $x$. We solve the
equations of motion of  the effective field theory within the WKB
approximation and find  the physical vacuum and the transitions
between the false (perturbative) and physical vacua.  Classical
solutions which dominate transitions between the false and
physical vacua are kinks that cannot be decomposed into
perturbative series over the powers of $\alpha_s$.  These kinks
lead to color inflation and the Bose-Einstein condensation of
quasiparticles. The account of the quantum fluctuations around the
WKB solution reveals the appearance of the "massless" particles--
"phonons".  It is explained that "phonons" are relevant for the
black disk behavior of cross sections of small $x$ processes. The
Bose-Einstein condensation of the ladders produces a color network
occupying a "macroscopic" longitudinal volume. We discuss briefly
the possible detection of new QCD effects. We outline albeit
briefly the relationship between the small $x$ hard QCD processes
and the coherent critical phenomena.

 \end{abstract}
\maketitle

\setcounter{page}{1}
\section{Introduction}

\par
One of the challenging properties of QCD is the rapid increase
with energy of the cross sections of  the hard processes .
Initially the increase has been predicted within the leading order
DGLAP approximation \cite{Gross,Dokshitser} . The rapid increase
of the structure functions of the proton with the energy has been
observed in ep scattering, for the review and proper references
see ref. \cite{AC}. The deep-inelastic structure functions of a
proton, calculated in the perturbative QCD (pQCD) within  the
leading twist (LT) approximation, can be fitted at small $x$ as
$F_{2p}(x,Q^2),xG_{p}(x,Q^2)\propto  x^{-\mu(Q^2)}$, where $\mu
\approx 0.2$ at $Q^2\approx 10$~GeV$^2$.  The generalization of
the QCD factorization theorems to the amplitudes of hard
diffractive processes shows that these cross sections (that are
higher-twist effects) should increase with energy even faster than
the structure functions, calculated in the leading-twist
approximation \cite{BGFMS94,CFS}. Predicted by the perturbative
QCD increase $\propto  x^{-2\mu(Q^2)}$ of cross sections of hard
diffractive processes, has been recently observed at HERA (DESY)
in the diffractive photoproduction of the heavy flavor mesons
$J/\psi,Y$,  in the high $Q^2$ diffractive electroproduction  of
$\rho^0$ mesons, for the review and references see ref. \cite{AC}.
The rapid increase of amplitudes of hard processes with energy
predicted in perturbative QCD within the DGLAP approximation  can
not continue forever since it violates at sufficiently small $x$
the strict inequality:  $\sigma(tot)\ge \sigma(diff)$. Here
$\sigma (tot)$ is the full,  and $\sigma(diff)$ is the
differential cross-section, and the cross-section $\sigma(tot)$ is
calculated within the LT approximation,  see ref. \cite{AFS}.

\par
To quantify the theoretical challenge it is useful to introduce an
auxiliary theoretical object - the amplitude describing the
scattering of colorless and spatially small dipole of the
transverse size $\approx 1/Q$ of a hadron target or of another
small size dipole. It follows from the probability conservation
that properly normalized such amplitude for the scattering at the
fixed impact parameter $b$ should be restricted from above :
$\vert f(b,Q^2,s)\vert \le 1$,  see refs. \cite{Strikman,Iancou}
and references therein. This inequality is violated within the
DGLAP approximation beyond the value
 of Bjorken $x$ equal to $x_{cr}(Q^2)$. Probability conservation restricts region of
 applicability of the DGLAP approximation but it does not precludes the fast increase of  the dipole cross section
$\sigma_{dipole T}\sim \log^2(s/s_0)$ corresponding to black disk
limit, cf. ref. \cite{Frankfurt}. The structure functions should
increase with energy even faster : $F_{2 p},xG_{p}\propto Q^2
\ln^3(x_o/x)$ resulting from the increase with energy of essential
impact parameters and from the ultraviolet divergence of hadronic
contribution into  renormalization of the electric charge, see
ref. \cite{Frankfurt}.

\par
The applicability of the competing leading order (LO)
$\alpha_s\ln(x_o/x)$ approximation requires rather large energies,
because the convergency of the perturbative QCD series at small
$x$ is rather slow,  (c.f. eq. (2.5) below).  The evaluation of
the next-to-leading order  effects found huge correction and
significant reduction of the increase of the amplitudes with
energy obtained within LO approximation \cite{BFKL}  due to the
necessity to restore the energy-momentum conservation
\cite{Ciafaloni,Lipatov-Fadin}. The further reduction of the
energy dependence of the amplitude has been found in ref.
\cite{Ciafaloni}  where the separation of scales characteristic
for hard and soft QCD processes has been performed and the running
of the coupling constant and the energy-momentum conservation were
taken into account. The practical conclusion is that almost in the
whole range of LHC energies the LO+NLO DGLAP and BFKL
approximations predict rather similar energy dependence of the
structure functions, cf. ref. \cite{Salam}. It seems now that the
challenge of violation of probability conservation becomes
important at lesser energies than the difference between the
different pQCD calculations reveals itself at small x.

\par
Small $x$ physics problem for the two scale processes has been
discussed recently in
 \cite{MV,JKLW,Mueller,McLerran,Kovchegov-Balitsky,Iancou} within the color glass condensate
 (CGC)
approach. The unitarity  of scattering matrix has not been
achieved within the made approximations \cite{KW}. (See however
the refs. \cite{Mclerran1} whose approach is similar to that used
in ref. \cite{Frankfurt} but different  from the mechanism
discussed in this paper.)

\par
The perturbative QCD calculation of the scattering of a
color-neutral two-gluon dipole of the size $\approx 1/Q$ off a
proton target violates probability conservation for $Q^2\approx
10$ GeV$^2 $ for the scattering at central impact parameters at
$x\le x_{\rm cr}(Q^2)\sim 10^{-5}$--$10^{-4}$, i.e. in the
kinematics typical for the large hadron collider (LHC)
\cite{Strikman}. The puzzle arises in the kinematics when few
gluons are produced only.  It  reveals itself in the gluon
structure function of a proton as the consequence of  the large gluon
density in the initial condition for QCD evolution and the rapid
increase with energy of the perturbative QCD amplitudes
\cite{Strikman}.  The violation of the leading twist approximation
should become important at the values of $x$ where the
perturbative QCD calculations in the leading twist approximation
are still more or less unambiguous. The scale of the hard
processes $Q^2\sim 10$~GeV$^2$ is chosen to guarantee the
smallness of the running invariant charge. Note that the existence
of the discussed above puzzle  in the amplitude for the scattering
of two spatially small dipoles investigated in this paper is
unrelated to the poorly understood physics of the quark
confinement  and the spontaneous broken chiral symmetry.

\par
With  the increase of the collision energy the higher-twist
effects blow up: higher is the twist of  the term -- more rapidly
it increases with the energy.  At sufficiently small $x$ the
perturbative QCD series have the form:
 $$\sum_{n=0}^{n=\infty} c_n(1/x)^{n\mu}$$
 where coefficients $c_n$ are rapidly increasing with
$n$ . The rapid increase of $c_n$ with $n$ is  known for a long
time from the study of the topology of the iterations of the
ladder diagrams \cite{VGribov-AMigdal}. The challenging problem is
to develop the unambiguous method of summing the divergent series.
One of the aims of this paper is the reconstruction of the
nonperturbative terms related to this divergence, the terms that
can not be decomposed into powers of $\alpha_s$.

\par
At sufficiently large energies, the hadron scattering at central
impact parameters becomes completely absorptive, as the consequence of the compositeness of the projectile and the
increase of the interaction with the energy, see ref. \cite{Zhalov}
and references therein. The FNAL data on the differential cross
sections of elastic $pp$ collisions shows that the $pp$ scattering at central impact parameters is black
 (cf. discussion in
refs.~\cite{Islam,MartiBlok}). This fact can be considered as an experimental hint of expected limiting
 behavior of hard processes
also at small $x$ where the interactions become strong.

\par
Another distinctive  property of  small x hard processes
is the rapid increase with energy of the longitudinal distances
 important in the scattering process. The coherence length
 $l_c$ in the total cross section of the deep inelastic scattering
 in the kinematics near the black disk limit is  increasing with energy
as $l_c\propto x^{\mu (Q^2)-1}$. This increase is somewhat less
rapid  than the one familiar from the analysis of the leading
twist approximation, see ref. \cite{BF2} and references therein.
Nevertheless, at current and future accelerators $l_c$
significantly exceeds the static radius of a hadron or nucleus. In
fact, $l_c$ becomes comparable with the electromagnetic radius of
the hydrogen atom in the kinematic region to be studied at the
Large Hadron Collider. As a result, a variety of observable new
coherent phenomena are  expected to appear in the small-$x$
processes \cite{AFS}. In addition it has been shown that in the
coordinate space the correlators of the currents evaluated within
both the DGLAP and BFKL approximations increase rapidly with the
distance \cite{BF1,BF2}.  Such increase is a necessary condition
for  the onset of the critical phenomena \cite{LL}. The
perturbative QCD produces branch points in the angular momentum
plane located at $j\ge 1$ -- i.e. in the region forbidden by the
causality and conservation of probability. Thus the theory has a
tachyon.  In this respect, certain similarity may exist between
the theory of small $x$ phenomena and the theory of strings in 26
dimensions where tachyon is also present in the perturbative
(nonphysical) vacuum, cf.  discussion in ref. \cite{Shatashvily}.
In these theories the account of the nonperturbative phenomenona
is necessary to find a true ground state.

\par
The well understood property of QCD is the strong dependence on
Bjorken $x$ of the piece of the quark-gluon  component of the wave
function of the virtual photon that  dominates the structure
function of a hadron target. At moderately small  $x$,  both soft
components in the wave function of the virtual photon and the
quark-gluon configurations,  where the constituents have large
relative transverse momenta, give comparable contributions, in the
spirit of the aligned jet model \cite{Bj,FS}. With the increase of
the collision energy this conspiracy disappears: the hard
configurations that occupy most of the phase volume in the photon
wave function begin to dominate  due to the rapid increase of the
hard cross sections with the energy. This corresponds to the
serious change of the $Q^2$ dependence of the structure functions
to the regime where $F_2(x,Q^2)\propto Q^2\ln(x_0/x)^3$, for the
review and references see ref. \cite{FSW}. In the new regime hard
pQCD phenomena  should dominate in the structure functions. The
cross section of the diffractive production of quark-gluon system
with large mass determines the triple ladder vertex which enters
our calculations. In order to simplify our task we restrict
ourselves to the kinematics where the wave function of the
longitudinally polarized photon $\gamma_L^*(Q^2)$  is dominated by
the configurations of the constituents with  large transverse
momenta and large invariant mass---the hard  perturbative QCD
analogue of the triple-reggeon limit. The account of this
phenomenon helps to evaluate the triple reggeon vertex near the
black disk limit  where the leading logarithmic approximations are
violated.

\par
A necessary kinematical condition of the applicability of our method is
 $x\le x_{\rm cr}(Q^2)\cdot 10^{-2}$.  Here $x_{\rm cr}(Q^2)$ can
be determined from the condition that  the contribution of one
ladder in dipole--dipole scattering at central impact parameters
is near the black disk limit in the leading twist approximation.
This inequality is just the condition for the existence of the
triple-Regge limit and it  is the main kinematical limitation of
our approach.

\par
The aim of this paper is to show that some of the challenges discussed above   can be met for one-scale hard small $x$
processes,  such as
 $\gamma_L^*(Q^2)+\gamma_L^*(Q^2)\to$~hadrons, $\gamma+\gamma\to Y+Y$,  etc.
We restrict ourselves by the consideration of the one-scale hard processes where the contribution into
 the amplitude of the QCD evolution with scale is suppressed. Consequently the factorization of the information
  on the wave functions of the projectile and target
from the interaction, similar to the one that occurs in the Regge
pole exchanges, becomes a reasonable approximation.

\par
The rapid increase of the perturbative QCD amplitudes with energy
leads to even more rapid increase of the role of the shadowing
effects.  It was found  long ago that the generalization of the
technology of calculations of  shadowing effects, encoded in the
Reggeon Field Theory, may appear useful for the description of the
small $x$ processes \cite{DDT}.   The generalization of the latter
approach, which accounts for any diffractive processes, leads to
the effective field theory (EFT). The evident advantage of the
EFT, as compared to models based on elastic eikonal approximation
(see i.e. ref. \cite{Maor}) is in the  possibility to account for
the whole variety of rescattering effects, and  to ensure the
energy -momentum conservation. In addition,  we will show in this
paper, that the transition to the black disk limit  (BDL) in hard
processes has a certain resemblance to a critical phenomena .

\par
Note that the  soft QCD contribution to the scattering processes is almost absent in the chosen
processes. This suppression is achieved  by the choice of the longitudinally polarized highly virtual
 photons as the projectile and the target.
Although the method developed in the paper is inapplicable beyond the one scale hard processes,
 the new QCD phenomena we found
in the paper may appear important for the many-scale
hard processes as well.

\par
To account for the coherence of the  high-energy processes
and  the rapid increase with energy of the amplitudes of the hard small-$x$ processes,
 we construct an effective field theory (EFT)
of the interacting perturbative colorless ladders.  The evident
advantage of this approach is the significant reduction of the
number of variables in the problem and the mapping of the
physics of the coherent processes into the framework resembling
the statistical models. (This approach is in  the spirit of the statistical models of critical
phenomena that account for the interactions
between the major modes only. The specifics of the physics of the
large longitudinal distances is included in the concept of the
quasiparticle.) The interaction between the quasiparticles, when
the amplitudes of the hard processes are near the unitarity limit,
can be easily evaluated within the WKB approximation. The use of
the WKB approximation  is justified by the smallness of the
running coupling constant in the perturbative QCD . We show that
account of the running coupling constant helps to establish the
dominance of  the triple-ladder vertex (see Appendix B). The
smallness of the multiladder vertices in pQCD implies that the
basic phenomena that characterizes the black disk limit  should be
insensitive to the restriction by the triple-ladder vertex. This
observation helps to fix the form of the Lagrangian of the EFT
near the unitarity limit.

\par
Since  $\mu\ge 0.2$ in pQCD, the quasiparticle of the EFT is a
tachyon:  the singularities in the complex angular-momentum plane
are located at $j-1=n\mu$ where $n=1,2,\ldots$.  Such a behavior
is in the evident  conflict with the  bound for the total
cross section of a small dipole--dipole scattering, that follows
from the causality and conservation of probability in QCD . The
causality and unitarity of the $S$ matrix, however, are not
necessarily valid in the perturbative QCD vacuum.  The rapid
increase of the amplitude with energy predicted by the
perturbative QCD approximations (i.e. the presence of the tachyon
in the EFT) reflects a rapid transition from the false,
perturbative vacuum to the nonperturbative, physical vacuum. To
visualize this phenomenon it is useful to consider physical
processes in coordinate space  \cite{BF2}. So these approximations
should fail to predict actual energy dependence of amplitudes of
physical processes which is nonperturbative phenomenon.

\par
We find in the paper that account of nonlinear phenomena leads to
serious restructuring of produced QCD states besides the onset of
black disk limit. With the Lagrangian of the EFT at hand, we may
identify the order parameter relevant for the critical phenomena
in small-$x$ processes: the order parameter is the condensate of
the quasiparticles, that are the pQCD ladders.  We find the new
nonperturbative phenomena in the regime of the strong interaction
with a small coupling constant, including the new nonperturbative
classical fields, whose effects are $\propto \exp(1/\alpha_s)$,
the color inflation due to the tunneling transitions,  the zero
modes in the dominant classical fields and related  massless
particles -``phonons''  in the EFT, and the formation of a color
network where the overlapping quasiparticles (ladders) do exchange
the colored constituents.

\par
The approach developed in this paper  differs from the EFT
suggested in ref.~\cite{Lipatov}. The major practical differences
are that we restrict ourselves by one scale hard processes only,
in the dominance in the WKB approximation of  the tunneling
transitions and the quantum fluctuations around them, while the
contribution of the quantum loops calculated near the vacuum of
the pQCD vacuum is negligible.  Another fundamental difference
from the approach of \cite{Lipatov} as well as from the Color
Glass Condensate approaches (cf. reviews
\cite{Mueller,McLerran,Iancou})   is in the account of the
important role played by  the diffusion to large impact
parameters.
 Near the unitarity limit this diffusion leads to the fast rise
of structure functions with energy \cite{Frankfurt}. In our paper
we take into account the universal Gribov diffusion related to
randomness of radiation \cite{Gribovdiff}, but neglect the
diffusion to large distances due to the running coupling constant
and the diffusion to small distances related to the increase of
the final state phase volume. The latter two types of diffusion
were shown in ref. \cite{Ciafaloni}  to be strongly   suppressed
after an account of the next-to-leading order BFKL type
approximation .  There is no systematic evaluation of these
effects because of the sensitivity to the restriction of the phase
volume of the produced particles and related sensitivity to NLO
approximations.

\par
In this paper we heavily use results of analysis of particular
preQCD model of Reggeon field theory obtained  in
\cite{amati1,amati2,amati3,amati4,amati5}. The major difference
from these papers is that dominance of triple ladder vertex is
justified in QCD .  Besides we found that EFT predicts significant
overlap between ladders near BDL . In the  Appendix A we account
for the exchange of the constituents between the overlapping
ladders. At sufficiently small $x$  this leads to the formation of
the color network instead of a system of ladders (see discussion
in Appendix A).

\par
The paper is organized in the following way. In Section~2 we
introduce the concept of the pQCD ladder as the quasiparticle
of the EFT.  We explain the hierarchy among the multiladder
vertices in pQCD (see also Appendix~B). We show that a
restriction to the triple-ladder vertex evaluated within the
physical vacuum  is sufficient  for the theoretical description of
the  phenomena near the unitarity transition,  and we construct the Lagrangian of the EFT.
 In Sections~3 and ~4 we discover
and analyze the analogies with the critical phenomena. In
Section~5 we briefly discuss observable phenomena that follow from
the solution of the effective field theory. Our results are
summarized in the conclusion. In Appendix~A we show that color is
confined within ladders in the tube of the small transverse size.
We argue that a color network appears once the ladders overlap
significantly in space. In Appendix~B we present the estimates of
the multiladder vertices within the physical vacuum.

\section{Effective field theory in QCD}

\par
The construction of an effective field theory requires the
knowledge of the dominant degrees of freedom---the quasiparticles---and their interactions. We assume, based on
 the linear pQCD calculations, that the  major degrees of freedom are the pQCD color-singlet
 ladders with the two-gluon state in the $t$ channel, that we
will denote the {\em "Pomeron"}. (The four gluon exchange , where each pair of gluons forms
an $8_F$ representation of the SU(3)$_c$, is included in this Pomeron.) Such a definition is
of common use, although the pQCD "Pomeron" has no direct relation to the Regge-pole Pomeron exchange
relevant for the successful description
of the phenomena dominated by the nonperturbative soft QCD. In principle, the intercept of "Pomeron"
 depends on $Q^2$ in pQCD.
 However at very small $x$ the $Q^2$ dependence of "Pomeron"
 intercept becomes weak \cite{Ciafaloni,Lipatov2}, and
 the factorization of information on projectile and target
 is valid for a ladder because of the disappearance of the color transparency phenomenon.

\par
The contribution of the pQCD ladders that  have 4, 6, $\ldots$
gluons in the $t$ channel where pairs of gluons form the color representations $8_{D}$, 10, $\bar{10}$, or 27 cannot
be topologically and analytically reduced to the contribution of
the two-gluon exchange ladder. In principle it is necessary to
introduce the new varieties of quasiparticles into the Hamiltonian
of the EFT and to account for their interactions. However, the smallness of the running coupling constant leads to the hierarchy
of the interactions especially within large $N_c$ limit evaluated
within the physical vacuum .  The
account of this hierarchy  justifies the neglect of this kind
 of quasiparticles. In particular, it is well known that the two-gluon
colorless ladder has maximal $\mu$ as
compared to the colored ladders.  The  contributions of all
quasiparticles except the ladders with the two gluon color singlet
 intermediate states in $t$ channel  can be neglected even near the unitarity limit, where the amplitudes of the high-energy processes
are close to the maximum permitted by the probability conservation.

\par
When the contribution of one ladder to the amplitude becomes large
it is necessary to take into account the shadowing effects. The
technology of the evaluation of the rescattering effects due to
the iterations of the pQCD ladder is the adaptation to pQCD of
methods of the Reggeon Field Theory(RFT) \cite{GribovCalc}. The
account of the shadowing effects leads to the 2+1 dimensional
field theory where variables are the position of the leading
singularity in the angular momentum plane--$j-1$ and the
transverse momentum $k_t$. The equivalent but more convenient
approach to the problem considered in the paper is to use the
description  in terms of the variables $y$ -rapidity and $\vec b$
--impact parameter. The effective Field Theory includes  the
"Pomeron" loops  and all variety of multi-"Pomeron" vertices.
Technologically it is more convenient to find the Lagrangian of
the EFT rather than to evaluate the particular set of diagrams,
cf. ref. \cite{VGribov-AMigdal}. Consequently, we may restrict
ourselves to the brief description of the Lagrangian of the EFT
near the unitarity limit.

The equations of the EFT  can be derived  as the Lagrange
equations of motion from an effective Lagrangian that contains
five terms, \beq L=L_0-L_1-L_2-L_3  -L_4. \label{1} \eeq The first
three terms have a straightforward interpretation in the case of
noninteracting  quasiparticles,  while the  other terms describe
interactions between quasiparticles and with the virtual photon as
the source. The $L_{0}$ , $L_{1}$ and ~$L_{2}$ follow from the
Mellin transformation of the Green function of the free
quasiparticle, $G=[j-1-\mu(Q^2)-\alpha' k^2]^{-1}$, in the plane
of the complex angular momentum $j$ in the crossed channel. Thus
\beq L_0=\frac{1}{2}(q\partial_y p -p\partial_y q), \label{2} \eeq
where $p(y,b)=\psi^+$ and $q(y,b)=\psi$ are the quasiparticle
fields. We denote $\partial_y=\partial_{\log(x_{0}/x)}$ where $y$
is rapidity and $x=Q^2/(2PQ)$, P is the 4-momentum of the target.
The quantity $x_0\approx 0.1$ denotes the length of the
fragmentation region where there are no $\log(x_0/x)$ factors.
$L_1$ is the ``mass'' term, \beq L_1=-\mu(Q^2) pq. \label{3} \eeq
This term accounts for the rise of the structure functions with
energy as derived for the scattering of two dipoles within the LT
approximation, \beq F_{2p}(x,Q^2),\ xG_{p}(x,Q^2)\sim (x_0/x)^{\mu
(Q^2)}, \eeq with $\mu>0$. The intercept $\mu(Q^2)$ has been
evaluated in the pQCD in the next-to-leading order (for a review
and references see ref.~\cite{Salam}), \beq \mu \approx 4\log
2\alpha_s(Q^2)N_c/\pi
(1-4(\alpha_sN_c/\pi)^{2/3}-6.5(\alpha_sN_c/\pi)+O(\alpha_s)^{4/3}+\cdots),
\label{t1} \eeq where $\alpha_s$ is the effective coupling
constant and $N_c$ is the number of colors. The dependence of $\mu
(Q^2)$ on $Q^2$ is rather weak at sufficiently small x
\cite{Salam,Lipatov2}. Note also slow convergence of pQCD effects.
\par
The term $L_2$ describes the dependence on  the collision energy
of the essential  impact parameters. We assume that it has the
form: \beq L_2=\alpha'_{P}p\triangle_{\vec b}q. \label{4} \eeq
Here $\vec b$ is a two-dimensional impact parameter, and
$\alpha_{P}'$ is the "Pomeron" slope, $\triangle_{\vec b}$ is
Laplace operator in "b" space. The slope $\alpha_{P}'$ is  small
within the perturbative QCD.  Within the approximation, that takes
into account the $Q^2$ evolution, the running of the coupling
constant, and to some extent the energy-momentum conservation,
the fast QCD evolution to small and large distances (that exists
beyond Gribov diffusion) is suppressed  \cite{Ciafaloni},
 and we can neglect it.  The Gribov diffusion to
large impact parameters  due to randomness of gluon radiation
\cite{Gribovdiff} is still there. The parametric estimate gives:
$\alpha_{P}'\propto N_c \alpha_s(Q^2)/Q^{2}$.   At the same time,
near the unitarity limit the additional  mechanism ( complementary
to Gribov diffusion) of increase with energy of essential impact
parameters,  due to the fast increase of the amplitudes with
energy, begins to play an important role
\cite{Frankfurt,Guzey,Strikman}.  There the effective
$\alpha_{P}'$ cannot be negligible but it is difficult to evaluate
it  because of the sensitivity to the uncalculated nonleading
order terms in the running coupling constant, etc. Since the exact
form of $L_2$ is  unknown at present it is chosen in the paper in
the simplest form.

\par
The evaluation of the multi-Pomeron vertices near the unitarity limit
 (see Appendix B) shows that within the WKB approximation, the relative contribution of the fourth and higher multi-Pomeron vertices
is suppressed by the powers of $\alpha_s$ compared to the
triple-ladder term. Thus for the description of hard QCD phenomena
it should be sufficient to restrict ourselves to the triple-ladder
interaction.

\par
Near the unitarity limit the dominant contribution into the triple-"Pomeron" vertex $\kappa$:
\beq
L_3=\kappa pq(p+q),
\label{11}
\eeq
is given by the component of the virtual photon wave function
having transverse size $\le 1/Q$.  Really at moderately small
$x$ soft QCD and pQCD components of the photon wave function  give comparable contributions into the total cross section and cross
section of the diffraction . This is due to the conspiracy of small
probability of the soft component $\propto 1/Q^2$ and the large
cross section of the soft component interaction with a target. The probability of a hard component  in the virtual photon wave function
is  $\approx 1$ but the cross section of its interaction with a hadron target is $\propto 1/Q^2$ - the aligned jet model \cite{Bj,FS}. This conspiracy disappears
at sufficiently small $x$ as the consequence of the rapid increase
with the energy of the cross sections of the hard processes. So small
$x$,  and therefore small $\kappa$ processes are dominated by hard QCD, for the review and references see ref. \cite{FSW}.
Then at moderately small $x$ the vertex $\kappa$ is determined by the interplay of the soft and hard QCD dynamics. However with the decrease of $x$ in the kinematics near the black disk limit the
vertex $\kappa$ should be dominated by the hard QCD.

\begin{figure}[htbp]
\centerline{\epsfig{figure=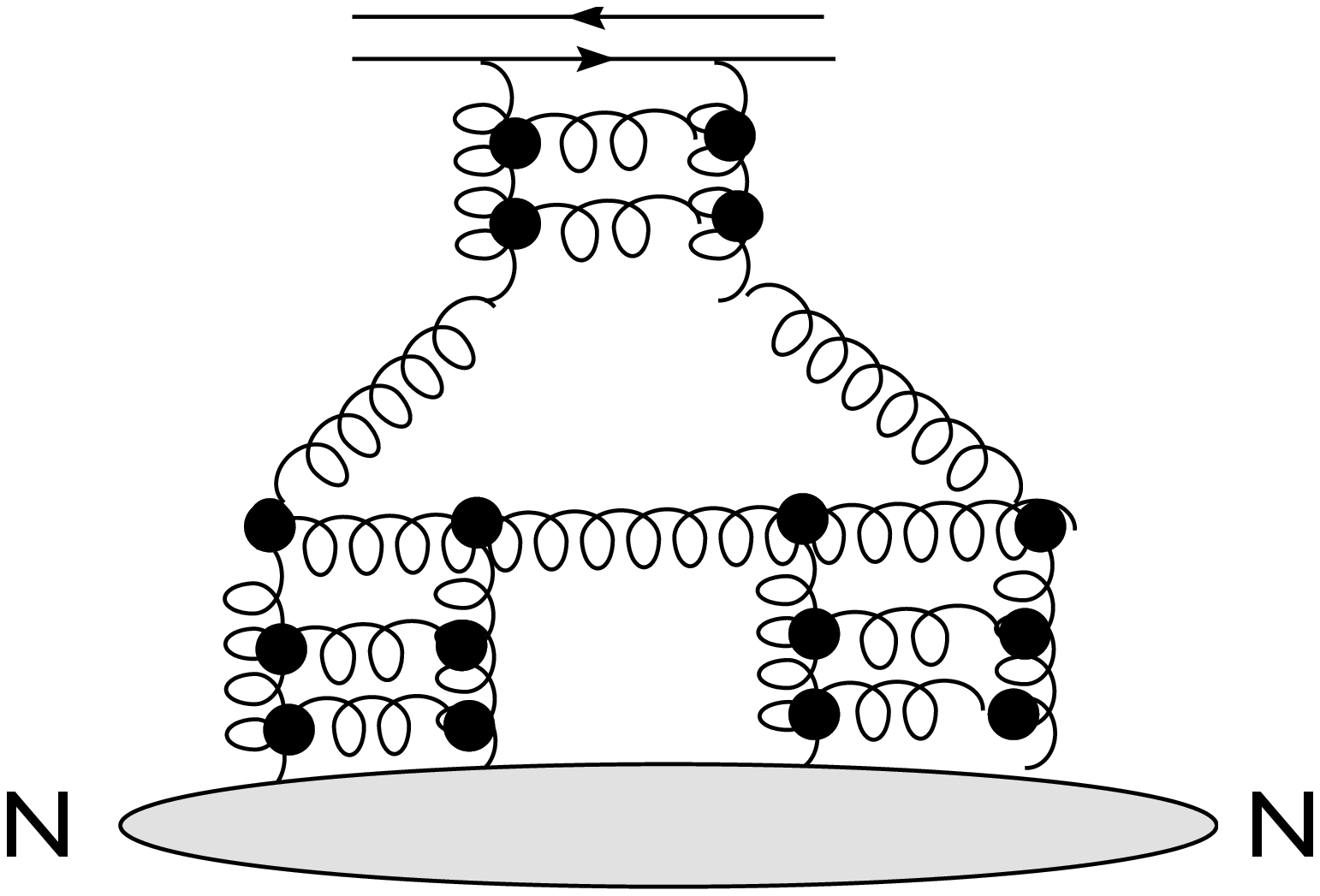,width=15cm,height=15cm,angle=-90,clip=}}
\caption{Three Reggeon vertex}
\label{SK}
\end{figure}

In the absence of a detailed calculation, we limit  ourselves to
the dimensional estimate of $\kappa$ and neglect any possible weak
dependence of the coupling $\kappa$ on $y$ and on $t$. In  the
lowest order in the coupling constant, the triple-"Pomeron"
vertex, is due to the interaction of ladders via one gluon loop
(see Fig.~\ref{SK}), where near the black disk limit hard QCD
dominates, as explained in the above discussion.  In addition the
account of the running coupling constant, and Sudakov form factors
suppresses the contribution of almost on shell effects . Our
estimate gives
\beq
\kappa \propto  \frac{\alpha_s^2 N_c}{\lambda} .
\label{14}
\eeq
Here $\lambda\propto Q$ and it is  somewhat increasing with the energy increase. This estimate of $\lambda$
differs from the one  obtained within the leading order BFKL
approximation far from the black disk limit.  The singularity in
the dependence of the triple-"Pomeron"  vertex on the momentum
transfer found in ref. \cite{Bartels} reveals an important role of
the soft QCD, the appearance of the effects encoded into aligned
jet model and represents another challenging problem for the BFKL
approximation at moderately small $x$.  We follow the
method of calculations of ref. \cite{Ciafaloni}  where the BFKL
approximation arises as the small $x$ limit of pQCD series where
separation of scales is made before summing them.
In this case for single scale hard processes $\kappa$ is dominated
by hard pQCD effects.  Besides the nonperturbative contribution
into the three-- "Pomeron" vertex is suppressed  by Sudakov form
factors, and by the dominance of the hard regime near the black
disk  limit, cf. the above discussion. The existence (but not the
properties) of the new QCD phenomena is not sensitive to the
actual  value of $\lambda $ which is effectively the
characteristic transverse momentum of the constituents of the
pQCD ladder where it splits into the two new ones.

\par
A tricky point in the evaluation of the "multi-Pomeron"
vertices within the pQCD is the necessity to  account
for causality and energy-momentum conservation. Diagrams
where singularities are located on the same side of the contour of integration in the energy plane (in particular eikonal diagrams) are suppressed by the powers of energy \cite{Mandelstam,GribovCalc}).
So we neglect the eikonal-type inelastic rescatterings since a bare particle may have one inelastic collision and any number of elastic collisions . For the interactions that rapidly increase with the energy, the requirements of the causality, the positivity of the probability for
the physical processes, and the energy-momentum conservation
can be hardly satisfied within such a set of diagrams \cite{FSW}.
In contrast, the contribution of rescatterings  due to an inelastic diffraction into the final state with the invariant mass $M^2$, where
$\beta=Q^2/(Q^2+M^2)$, is not too small.  This contribution
dominates in the two-scale hard small-$x$ phenomena
at $x\approx x_{cr}$. We include this contribution in the scale factor of the source. (This contribution can be interpreted as the coupling of
the ``Pomeron'' to secondary reggeon trajectories.)

\par
Within the approximations made in this paper the coupling of the pQCD ladder to a hadron can be treated as  the interaction with a source . The actual form of the  $L_{4}\equiv L_{\gamma}$ term is unimportant for the most of the results obtained in this paper.

Nonetheless, let us discuss the interaction of the ladders with a
virtual photon as an external source in the center-of-mass frame
of the reaction.  There is the coupling with the projectile and target virtual photons:
\beq
L_\gamma =\int_0^1 \,\frac {dz}{z}\,d^2b\, \Phi^2(z,\vec b,Q^2)
q(z,\vec B-\vec b) c_{\rm dipole}\delta(Y+y).
\label{la}
\eeq
Here $Y$ is the total rapidity, and $\vec B$ is the total impact parameter. In this formula we neglect the $Q^2$ dependence of
the interaction in the vicinity of the black disk limit because, near
the black disk limit, the color transparency phenomena and the
related decomposition over twists disappear. In the above
equation $\Phi^2(z,\vec b,Q^2)$ is the square of the dipole wave
 function of the virtual photon in the transverse parameter space,
$z$ is the fraction of the photon momentum carried by the
constituent, and  $dz/z=dy$, where $y$ is the rapidity. The
variable $\vec b $ is  the transverse distance between the
constituents in the photon wave function. The dipole wave function
is given by (for simplicity we restrict ourselves to the
collisions of longitudinally polarized photons only) \beq
\Phi_L(b,z,Q^2)^2=c(N_c\alpha_{\rm em})(Qz(1-z))^2)
K_0^2(bQ\sqrt{z(1-z)}). \label{dipole} \eeq The numerical factor
$c$ follows from a convention on the normalization of this wave
function. Note that this function is localized in the space of
relative transverse distances between the constituents as
$\exp(-bQ\sqrt{z(1-z)})$ and in the space of rapidities $y$. Since
the distribution over $z$ is centered at $z=1/2$ in the case of
the wave function of the longitudinally polarized photon, we can
safely assume that this contribution into the source corresponds
to the finite $y$ negligible as compared to the full rapidity
span. So we assume for certainty that $Y-y\sim 0$. The
contribution of inelastic diffraction that  dominates at not
extremely small $x$  \cite{FSW} corresponds to $-Y\le y \le
-Y+y_{\rm dif}$.  Since our interest is  in  $y\gg y_{dif}$ we
may skip in the analysis of EFT  the difference between $-Y+y_{\rm
dif}$ and $-Y$, although this approximation restricts region of
applicability of approximations made in the paper.

Altogether, we obtain the Lagrangian:
\begin{eqnarray}
L&=&1/2(q\partial_y p -p\partial_y q) -\alpha'p\triangle_b q+
\mu pq-\kappa pq(p+q)\nonumber\\[10pt]
&-& c_{\rm dipole}\int \exp(-bQ/2) q(y,\vec B-\vec b)d^2 b \delta(y+Y)\nonumber\\[10pt]
&-&c_{\rm dipole}\int \exp(-bQ/2) p(y,\vec B-\vec b)d^2b \delta(y -Y),
\nonumber \\[10pt]
\label{lag}
\end{eqnarray}
where $\kappa\propto (\alpha_s^2N_c/ \lambda)$, and
$c_{\rm dipole}$  accounts for the normalization
of the virtual photon wave function.

The interaction between overlapping ladders due to the exchange by
the constituents will be considered in the end of the paper.

\section{Critical phenomena in hard perturbative QCD regime
near the unitarity limit.}

\par
The form of the effective Lagrangian (\ref{lag}) relevant for the
single scale hard QCD phenomena near the unitarity limit is rather
close to preQCD model  analyzed in refs.
\cite{amati1,amati2,amati3,amati4,amati5} (except for the form of the source term ) . In  the further analysis we use the WKB solution of the Lagrangian equations of motion and quantum fluctuations around them  found in refs. \cite{amati1,amati2,amati3,amati4,amati5}  and adjust it to describe the hard QCD phenomena .

\subsection{Classical solutions of EFT.}

\par The distinctive property  of the Lagrangian (\ref{lag}) is the
existence,  in addition to  the usual perturbative vacua \beq
p=0,q=0, \label{v0} \eeq
 , of the two new vacua:
 \beq
p=\mu /\kappa, q=0 \label{v1} \eeq and \beq p=0, q=\mu /\kappa .
\label{v2} \eeq Since Lagrangian of EFT  is nonhermitean it has no
vacuum in the usual sense  and  the term "vacuum" really means the
critical points of the action (\ref{lag}) of the EFT. The action
(\ref{lag}) has actually four critical points: the three critical
points (\ref{v0}),(\ref{v1}) and an additional critical point
$(\mu/(3\kappa),\mu/(3\kappa))$. However the contribution of the
last point to the S-matrix (see below) is suppressed by
$\exp(-HY)\sim \exp(-(\mu^3/(27\kappa^2)Y)$ relative to three
critical points (\ref{v0}),(\ref{v1}). Consequently the fourth
critical point can be neglected.
\par
Having Lagrangian (\ref{lag})  we may deduce the Lagrangian
equations of motion that will be analyzed first in the classical
approximation. The system of nonlinear partial differential equations
in 2+1 dimensions is
\begin{eqnarray}
-dq/dy&=&\alpha'\triangle q-\mu q +\kappa (2pq+q^2)\nonumber\\[10pt]
dp/dy&=&\alpha'\triangle p-\mu p+\kappa (2pq+p^2).\nonumber\\[10pt]
\label{v4}
\end{eqnarray}
This system has kink solutions. The boundary conditions in
the system (\ref{v4}) are at $y=\pm \infty$ and
correspond to the vacua (\ref{v0}),(\ref{v1}),(\ref{v2}).

\par
The detailed analysis of the kink solution is possible in 1+1
dimensions only, where the above equations can be reduced to
the ordinary differential equation.  In refs. \cite{amati1,amati2,amati3} the family of kinks, characterized by a 2d velocity parameter v has been found. These kinks  interpolate between two vacua, say
(\ref{v1}) and (\ref{v2}). The action of the kink is finite
$S\sim (\mu/\kappa)^2(2-v)\phi_0^2$, where $\phi_0$ is the field
value at the impact parameter value $b=vY$, and $v $  is the kink velocity.  It  is proportional to $1/\alpha_s^2$, where we used the dependence of $\mu$ and $\kappa$ on $N_c$ discussed in
section 2.  For the value of parameter $v=2$ we obtain critical
kinks with zero action.  Quantum fluctuations around these kinks
are described by a positive quadratic form, cf. \cite{amati3}.

\par
The characteristic property of the kinks, both in unphysical 1+1,
and in physical 2+1 dimensions,  is their step function form.
One of the functions p or q behaves like a step function
\beq
p (q)\sim \theta (v\sqrt{\alpha'\mu} (y-y_0)-\vert \vec b-\vec b_0\vert),
\label{cf}
\eeq
where v is the kink velocity (a free parameter, v=2 for critical kink).
The solution contains an arbitrary parameter $y_0$ that helps to understand why the physics related to the fragmentation
can be hidden into the properties of the source. The arbitrary
solution depends also on $\vec b-\vec b_0$. The value
of $\vec b_0$ is not fixed by the equations. This is a zero mode relevant for the appearance of the "phonons" in quantum fluctuations.

\par
The calculations in  2+1 dimensions are significantly more
difficult. However one can easily prove the existence of the
kinks with the finite action that tunnel between the vacua
(\ref{v0}) and (\ref{v1}). We need to find a
solution of eq. (\ref{v4}) with boundary conditions
\begin{eqnarray}
y&\rightarrow&\infty\,\,\,p\rightarrow \mu/\kappa\,\,
q\rightarrow 0\nonumber\\[10pt]
y&\rightarrow& -\infty,\,\, q,p\rightarrow
0.\nonumber\\[10pt]
\label{v4a}
\end{eqnarray}
Then for $y\rightarrow -\infty$ it is legitimate to neglect by
nonlinear interaction and obtain the usual diffusion equations:
\begin{eqnarray}
-dq/dy&=&\alpha'\triangle q-\mu q \nonumber\\[10pt]
dp/dy&=&\alpha'\triangle p-\mu p.\nonumber\\[10pt]
\label{v6}
\end{eqnarray}
The general solution, say for q, is given by
\beq
q(y,b)=\exp(\mu y)\int \exp(-((\vec b-\vec b')^2/4\alpha_{P}'y)) f(\vec b')d^2b.'
\label{v7}
\eeq
Here $f(b)=cq(y=0,b)$ where c is a  numerical factor.

\par
The boundary condition for p will be $p\rightarrow \mu/\kappa$ for
$y\rightarrow \infty$. Then near $y\rightarrow\infty$, we may
write $p=\mu/\kappa -z$,  where z is small. The linearized
equation for z is \beq dz/dy=\alpha'\triangle z-\mu z+2\mu q
\label{z} \eeq The system of equations (\ref{v4}) imposes the
condition $q\rightarrow 0$ for $y\rightarrow \infty$ , since
$\alpha' \triangle p\sim 2\mu q$, and $p\rightarrow$ constant.
Thus for each value of the impact parameter $\vec b$ the function
 q has a maximum. It follows from eq. (\ref{z}) that  when z  is small (and y is approaching  $Y$), the field p decreases sharply near $Y$, so it can be approximated by the step function, while $q$ reaches
 its maximum at $y\sim Y$ and then goes to zero.

\par
It is easy to see that the action of the kink is finite and
proportional to $(\mu/\kappa)$ in some power.

\par
This solution is similar to the one found in the 1+1 dimensional
theory. However there is no analytical expression for the solution giving critical kinks with zero action as well as a full classification of kinks .  So in 2+1 dimensions we rely on the results of the
numerical simulations  made in refs. \cite{amati1,amati2},  who found the same properties of kinks for the 2+1 dimensional theory as
for 1+1 dimensional one.

\par
The existence of the step like solutions made it possible
to calculate S-matrix for the case of the sources given by eq.(\ref{dipole}). Indeed,  the classical equations
of motion for this case have the form
\begin{eqnarray}
-dp/dy&=&\alpha'\triangle p-\mu p +\kappa (2pq+qp^2)+
c_{\rm dipole}\int d^2 B \exp(-BQ/2)q(y,b-B) \delta (y-Y)\nonumber\\[10pt]
dq/dy&=&\alpha'\triangle q-\mu q+\kappa (2pq+q^2)+
c_{\rm dipole}\int d^2B \exp(-BQ/2)p(y,b-B) \delta (y+Y).\nonumber\\[10pt]
\label{v41}
\end{eqnarray}
Since coupling to dipole for the field q is much smaller than
$\mu/\kappa\sim 1/N_R^3\alpha_s^2$  these equations are
close to that analyzed in ref. \cite{amati1}. The field p is  small
for $y\le 0$, and it is
the solution of the usual diffusion equation  with the boundary
condition that follows from the first of eqs. (\ref{v41}), times
$\exp(\mu y)$.The field  p starts to rise exponentially for
$y\sim 1/\mu$. At this point one cannot however neglect the
nonlinear terms. Since for $y\rightarrow \infty$ $p\rightarrow
\mu/\kappa$, we can linearize eqs. (\ref{v4}) as $p=\mu/\kappa
-z$. Then for z one gets equation
\beq
dz/dY=\alpha'\triangle z+\mu z.
\label{v8}
\eeq
The equation for z is the same equation as above,  except that
the mass term changes sign, and it is small up to $y\sim 1/\mu$,
when it blows up, leading to a step like decrease of p.

\par
The point where solutions for p  blows up can be estimated from the equation
\beq
q(b,y)\propto \exp(\mu y-b^2/(R^2+4\alpha_{P}'y)).
\label{v10}
\eeq
The transition occurs near the point where $q\sim 1$, i.e.
\beq
(R^2+4\alpha'y)\mu y\sim b^2.
\label{v11}
 \eeq
 Here $R\sim 1/\lambda$ is the transverse scale in the problem.
 Asymptotically
 \beq
p(b,y)=(\mu/\kappa)\theta (2\sqrt{\alpha'\mu}y+\delta-b),
\label{as}
\eeq
where $\delta$ is a phase shift determined by boundary conditions.

\par
We see that in the classical approximation the equations for p and q effectively decouple. Moreover it can be proved that in the classical limit it is enough to solve the decoupled equations with suitably
chosen boundary conditions.

\par
The classical solution leads to the black disk limit, since in the
classical theory the S-matrix is given by \beq S\sim \exp(-\int
 d^2 b g(b)p(b,Y)), \label{rse} \eeq cf. refs.
\cite{amati1,amati2}. Here $g(b)$ characterizes the photon and it
is concentrated near $b\sim B$. Since $p(b,Y)$ has a step function
form, and  g(b) is concentrated near B, in the classical theory
S-matrix is 1 outside the black disk and zero inside (total
absorption).

\subsection{ Quantum fluctuations around the WKB solutions.}

\par
The knowledge of the family of the kink solutions permits the
semiclassical quantization of the theory and the calculation of
the  S-matrix.

\par
The distinctive feature of the semiclassical approximation is the
existence of the "critical" kinks, with zero action, in
particular, the kinks with both the energy and the classical
momentum being zero. Classical contributions of these kinks into
wave function are not exponentially suppressed. (Energy and
momentum of kinks are defined as components of $\int d^2b T^{00}$
and $\int d^2b T^{0i}$, where  the energy momentum tensor of the
EFT is found via the Nether theorem).

\par
The remarkable property of the quantum fluctuations around
critical kinks is the existence  of the zero modes-"phonons" in
the EFT , that are characterized by the linear dispersion :
\beq
E=i2\sqrt{\alpha'\mu}k=2\sqrt{\alpha'\mu} P_{\rm cl},
\label{C1}
\eeq
where $P_{\rm cl}=\int d^2b p\frac{dq}{db}$ is the total
classical momentum derived from the EFT action. We present here
plausible reasoning for the existence of such modes, following
refs.\cite{amati4,amati5}. Let us begin from the analysis of 1+1
dimensional model. In this case kinks are  f(x-vY) solutions of
the ordinary differential equations. Quantum fluctuations around
kinks can be calculated using the standard procedure
\cite{thooft}. The zero mode (\ref{C1}) leads to the gapless
spectrum of excitations ("phonons"). The "phonon" wave function
satisfies the equation:
\beq
H_R\vert k>_{\pm}=\pm i2\sqrt{\alpha'\mu}k\vert k>_{\pm}
\label{C2}
\eeq
and
\beq P_R\vert k>_{\pm}=k\vert k>_{\pm},
\label{C3}
\eeq
where $H_R$ and $P_R$ are the total  Hamiltonian and momentum
of the quantized EFT ($P_R=\int db T^{01}$ and $H_R=\int db
T^{00}$, where $T^{00}, T^{01}$ were calculated via the Nether
theorem). The corresponding full set of wave functions can be
found explicitly:
\beq
\vert k>_{\pm}=\int da exp(-ika)(\exp(-(\mu/\kappa)
\phi(\pm\rho-a)q(\rho)) \vert\Phi_0>.
\label{C4}
\eeq
Here $\rho=x-v_0Y$, and $\phi(\rho)$ is a solution
of a classical equation of motion for a kink, $\Phi_0$ is wave
function of a perturbative vacuum. These states propagate in one
direction in time $-$ to $p\rightarrow \mu/\kappa$ at
$Y\rightarrow \infty$ and in the opposite directions in impact
parameter space. It is straightforward  to calculate the
quasiparticle correlation function:
\beq
D(B,Y)=<\Phi_0\vert p(0,\vec 0)\exp(i\vec P\vec B)\exp(-HY)
q(0,\vec 0)\vert\Phi_0>,
\label{CF}
\eeq
using the set of states (\ref{C4}) together with a
perturbative and nonperturbative vacua, and get a black disk limit
as the asymptotic answer .

\par
Let us emphasize the key role of a linear spectrum. Indeed, for
the linear spectrum, as it was shown in ref. \cite{amati3},
\beq
D(B,Y)=\frac{\mu^2}{\kappa^2}  \int\frac{dk}{2\pi} \exp(ikB)
 \exp(iv_0kY)<\Phi_0\vert p(0)\vert k> <k\vert q(0)\vert\Phi_0>.
 \label{D1}
\eeq
It is possible to prove that the above matrix elements are equal to
\beq
<\Phi_0\vert p(0)\vert k>=-\frac{\mu}{\kappa}\frac{1}{\epsilon +ik}, \label{D2}
\eeq
where $\epsilon$ defines contour of integration around
singularities. Then we obtain:
 \beq
D(B,Y)=(\mu/\kappa)^2\theta(-B+2\sqrt{\alpha'\mu} Y).
\label{D3}
\eeq
In order to connect this correlator with the amplitudes we
have to  normalize it properly \cite{amati3,amati5}:
\beq
G(B,Y)=\frac{\kappa^2}{\mu^2}D(B,Y).
\label{ren}
\eeq
Then
 \beq
G(k,J)=\frac{2}{(J-1)^2+v_0^2k^2}
\label{D4}
\eeq
Here $v_0=2\sqrt{\alpha'\mu}$ is a critical kink speed. The scattering
amplitude is (\cite{GribovCalc}, Chapter 16)
\beq
A(s,t)=s\int^{i\infty}_{-i\infty}\frac{d\omega}{2\pi^2}
\exp(\omega\xi)G(\omega,k^2)(i+\tanh(\frac{\pi}{2}\omega))
\label{ast}
\eeq
Here
$\xi=\log(s), \omega=J-1, t=-k_t^2$.

\par
In addition, there exists a band of low lying states with a
dispersion relation $E\sim k^2$, and a band of states with a gap
$\sim \mu$. It can be argued that the first set of states
corresponds to the unshifted quasiparticle fluctuations around the
perturbative vacuum $\vert \Phi_0>$ in the presence of a kink and
viewed from the reference frame moving with a critical speed
$v_0$. The higher modes can be interpreted as the collective
fluctuations of a condensate of ladders, i.e. the quasiparticles
interacting with the kinks. It is easy to prove that these modes
do not influence the expanding disk solution asymptotically,
although can be important outside the disk, and near the
transition to a black disk regime.

\par
The numerical analysis of the spectrum of the quantum fluctuations
around the kinks has been performed  in refs.
\cite{amati1,amati2}. Moreover, it was proved on the discrete
lattice that the  Field Theory can be continuously connected with
the Ising model in transverse magnetic field. Two types of
collective excitations were found--the zero modes with a spectrum
$E\sim k$ and solutions with a gap and spectrum $E\sim a+bk^2$,
similar to the ones found in the 1+1 dimensional model in the
continuum. Evidently, only solutions with linear spectrum are
relevant for the asymptotic behavior of the high energy processes.

\par
To summarize, the quasiclassical solution of EFT has following
distinctive features:

\par A)
EFT has three  degenerated "vacua" (\ref{v0}),  (\ref{v1}), (\ref{v2}) (The word  "vacua" is in the brackets, since EFT is the theory
with nonhermitian Hamiltonian, and actually means critical points
of the action with zero value of the Hamiltonian). The true wave
function of the physical ground state is a linear combination of
these three vacua, and the Hamiltonian is diagonalized by
"critical" kinks with zero action. If the initial state is a
perturbative one interacting with a source, it evolves
 in rapidity and becomes a condensate of quasiparticles (ladders).
(The term "ground state" means here that all states in
the relevant physical sector the theory are the excitations of
this state).

\par B)
The S-matrix is given by the functional integral \beq
S(B,Y,g,f)=\int dp\int dq \exp(-L(B,Y,g,f)), \label{v11a} \eeq
where the corresponding action is calculated via the classical
kink solution described above, but includes now the source
terms--the couplings with the virtual photons. The main
contribution to the S-matrix comes from the quantum fluctuations
around the critical kinks with the zero action. The spectrum of
the excitations begins from the massless excitations- "phonons".
The contribution of the kinks into the two- quasiparticle Green
functions that determines the cross-section is:
\beq
G(B,Y)=\theta(B^2-4\alpha'_P\mu Y^2)
\label{v12}
\eeq
for large Y, $\mu Y>> 1$.  This Green function defined by
eq.(\ref{ren}) is related in a usual way to a scattering amplitude
by the Mellin transformation, (see e.g. ref. \cite{GribovCalc},
chapter 16 and eq. (\ref{ast})).  We denote the perturbative
vacuum (\ref{v0})  as $\Phi_0$  .

\par
C) The obtained state is described by the asymptotic wave function
\beq \Psi(y)=\exp(-\frac{\mu}{\kappa}\int d^2bq(b,y)
\theta(b^2-4\alpha'\mu Y^2))\vert \Phi_0>. \label{slon3} \eeq Here
$\vert\Phi_0>$ is a perturbative vacuum. To the extent that the
correlations may be neglected the asymptotic vector (\ref{slon3})
is a coherent state \cite{amati3} and the S-matrix is given by
\beq
S(B,Y)=\exp(-\frac{\mu}{\kappa}\theta(2\sqrt{\alpha'\mu}Y-B)),
\label{slon31} \eeq where we explore that the target is localized
near the impact parameter $b\sim B$. The S-matrix as given by eq.
(\ref{slon31}) is 1 inside the black disk and  suppressed
exponentially as $\mu/\kappa \sim \sim 1/\alpha_s$, outside.  This
is just the BDL behavior. Note the nonperturbative structure of
eq. (\ref{v12}), and that the wave function $\Psi(y)$ can not be
obtained by decomposition over powers of $\alpha_s$.

\par
 D)  In other words in the limit of the infinite energies
$Y\rightarrow \infty$  the produced state corresponds to a
Bose-Einstein condensate of  ladders  in the entire space.
However,for finite energies the solution is the black disk of
radius $R^2=R_0^2+4\alpha'_P\mu Y^2$. The equation (\ref{slon3})
gives the exact form of the wave function of the Bose-Einstein
condensate of ladders as a function of rapidity Y.

\par
E) The crucial reason why  both quantum and classical
approximations lead to the black disk behavior is  the existence
of the "phonon" zero mode in the quantum kink Hamiltonian, that
has been proved on the lattice, and in the 1+1 dimensional case.
Moreover,the analysis of refs. \cite{amati1,amati2,amati3} shows
that the contribution of the excited states that are different
from the "phonons" is relevant only in the vicinity of the
transition point. There are two types of states: collective
fluctuations near the perturbative vacuum interacting with quantum
kinks and the quantum fluctuations of the kink condensate. None of
them influences the asymptotic behavior of the black disk.

\section{Kinks and QCD}

Some properties of the classical solutions of the effective theory
can be understood directly in QCD. A kink produces action
proportional to $(\mu/\kappa )\sim (1 /\alpha_s)$ .  The
dependence  of the S-matrix (\ref{slon31}), of the critical kink
action (understood as the limit of a family of kinks with nonzero
action)  on the coupling constant $\alpha_s$  and existence of "phonon" show that this is a novel nonperturbative QCD
phenomenon.

\par
Let us note here that the 2d translational invariance is broken in
the theory at finite rapidity Y due to the existence of black disk
behavior. Mathematically it reveals itself in the existence of a
translational zero mode for the kink solution. This phenomenon
resembles a spontaneous symmetry breaking since the system, after
being at moderate rapidities in the perturbative vacuum chooses in
the process of space-time evolution the state which is a definite
combination of the vacuum states of the system. Whether this
phenomenon corresponds to the conventional  spontaneous  symmetry
breaking is unclear at present but the existence of "phonons"
looks suggestive.  Moreover  it is easy to show that in the
approximation where $\alpha'=0$  EFT can be mapped by substitution
of variables into the quantum mechanics with hermitian double well
potential with the two nonperturbative critical points of the
action (\ref{lag}) corresponding to two minima of Higgs type
potential. In this case two nonperturbative vacuums of section 3
are real minima of the action and tunneling transitions are
relevant for the change of symmetry.

\par
The estimate $Y_c$  where  transition to the regime of
Bose-Einstein condensation of ladders follows from the
inequality $Y_c\ge \ln (10^{2}/ x_{cr})$ where $x_{\rm cr}$ can be  found from the requirement of conservation of probability,
 of unitarity of the S matrix cf. ref. \cite{Frankfurt}.
 (Additional factor $\le 10^{2}$ accounts  larger energies needed for
the applicability of triple "reggeon"=ladders limit and all
ladders should be near unitarity limit.)  Assuming $x_{cr}\sim
10^{-4}-10^{-5}$ cf. \cite{Frankfurt} we obtain  $Y_c\ge 14-16 $ .
At these rapidities the Bose-Einstein condensation of the ladders  starts.

\par
The characteristic form of the kink is the step-function in the Y
space, with the width of the order $\delta Y\approx \log (\delta
E/Q) =1/\mu$. The coherent length relevant for the evolution of
the kink is enhanced due to the large Lorentz slowing down
gamma-factor as \beq T_I \sim \delta E/Q^2\approx \exp
(1/\mu)/Q\approx 10^2/Q
 \label{v^*1}
 \eeq
In other words, for sufficiently low x we have $T_I\ll T_c$, where
$T_c$ is coherence length $T_c\sim 1/(Q x^{1-\mu})$. This rapid
transition  to the black disk regime  can be called the "color
inflation": one ladder due to the tunneling transition blows up
during time $T_I$ and creates an entire region of space filled
with the gluon ladders. During time $T_I$  approximately
$(\mu^2/\kappa^2)R(Y)^2$ ladders are created , where $R(Y)$ is a
black disk radius for a given rapidity $Y$.

\par
It is easy to evaluate the density of ladders in the coordinate
space by solving discussed above diffusion equations (cf. similar
analysis in ref. \cite{amati5}):
\beq
\vert \Psi
(y)>=\exp(-(\mu/\kappa)\int^{R(Y)}_0 d^2b q(b,Y)).
\label{vec1}
\eeq
The conjugate state is
\beq
<\Psi(y)\vert =\exp(+(\mu/\kappa)\int^{R(Y)}_0 d^2b p(b,Y)).
\label{vec11}
\eeq
Expanding these  states into series of powers over $q$ we obtain
the multiladder wave functions $\Psi(b_1,....b_N)$ ,
\beq
\Psi=\sum^{N=\infty}_{N=0}\int \Pi^{i=N}_{i=0}\int
d^2b_1..d^2b_i..d^2b_N \Psi (b_1,....b_N)q(b_1)......q(b_N)/(N!),
\label{psi}
\eeq
where $\Psi(b_1,...b_N)$ are products of theta
functions, limiting all the integration in eq. (\ref{psi}) inside
the black disk. We neglect the correlations (see section 3).

\par
The average distance between the ladders $l_t$ can be estimated
from the wave function of the condensate . Indeed, \beq
l_t^2=\frac{\int^{R(Y)}_0(\vec b_1-\vec
b_2)^2\Psi(b_1,b_2,...)\Psi^*(b_1,b_2,....)d^2b_1...d^2b_N..}
{(\int^{R(Y)}_0\Psi^*\Psi d^2b_1...)} \label{vec} \eeq The
integration is over the impact parameters of the ladders. The
condensate wave function is homogeneous inside the black disk .
Therefore \beq <p(b_1)....p(b_N)q(b_1)....q(b_N)>=1 \label{dddeer}
\eeq Consequently, the transverse distance between the ladders is
\beq l_t\sim \kappa/\mu\sim \alpha_s /\lambda \label{v*} \eeq On
the other hand,  the characteristic scale $d_t$ of a ladder in the
transverse parameter space is determined by  the coefficient in
the effective Lagrangian in front of the kinetic term, that is
equal to  $\sim N_c \alpha_s/\lambda^2$, hence \beq d_t^2 \sim N_c
\alpha_s/\lambda ^2 \label{v**} \eeq Consequently, \beq
l_t^2/d_t^2\approx \alpha_s/ N_{c} \ll 1. \label{v***} \eeq It
follows from the above estimate that the pQCD ladders overlap
significantly .  But overlapping ladders can exchange the quarks
and the gluons because in the perturbative QCD there are no
barriers between the ladders. The resulting color network, as we
shall call this object, is macroscopical in the longitudinal
direction. The distinctive features of the color network resemble
the quark-gluon plasma .

\par
Understanding  the actual longitudinal structure of the system is
not possible within the effective field theory. Indeed, even after
the phase transition, the ladders continue to grow till the color
network achieves the longitudinal length $1/Qx^{1-\mu}$.

\par
In our analysis we used the fact the QCD ladders can be considered
as the effective degrees of freedom up to the scales when they
significantly overlap . This overlap is controlled by the density
of ladders since in the perturbative QCD (see Appendix A) there
are no long range forces operating between the ladders.

\section{Observable phenomena.}

We predict a variety of the new nonperturbative QCD hard phenomena
for the case of the collision of the two small dipoles near the
black disk limit. Their very existence shows that transition to
the black disk limit is a kind of a critical phenomenon.

The phenomenon of  color inflation may significantly change the
physics of hard processes at sufficiently small x  by softening
the parton distribution over the longitudinal momenta and will
reveal itself as the threshold like increase of  multiplicity. (We
postpone quantitative analysis of such phenomena to the next
publications).

Comparatively clean way to identify the onset of the new regime
would be to measure Mueller-Navelet process \cite{MN}: $p+p\to
jet+X+jet$ where the distance in the rapidity between the high
$p_t$ jets is large. The expected behavior is the following:
initial fast increase of cross section with $y$ predicted by the
pQCD should change to the fast decrease at larger $y$ because of
the color inflation, i.e. disappearance of the long range
correlations in the rapidity (coordinate) space near the black
disk limit due to the creation of the color network.

\section{Conclusion}

\par
We argue that the absence of the long range forces between the
colorless ladders as the consequence of the color screening
phenomenon (see Appendix A) justifies the neglect of the
exchanges of the constituents between the colorless ladders in the
first approximation. The colorless ladders should be a dominant degree of freedom even near the black disk limit. As a result it is
possible to build the effective field theory where ladders are
quasiparticles.  We showed that at sufficiently small x  the form
of the Hamiltonian of this EFT is dictated for the single scale
hard processes in QCD by the smallness of running coupling
constant . Moreover smallness of the effective triple ladder vertex:
$\lambda^2 \kappa \ll 1$ justifies the applicability of
the semiclassical approximation. We show that the effective field
theory  is solvable within the WKB approximation and leads to the
black disk behavior  and other QCD phenomena.

\par
The transition to the black disk regime within the effective field
theory in one scale hard processes is a kind of a critical
phenomena. There exist classical hard QCD  fields relevant for
the transition from false to physical vacuum -kinks in the rapidity-impact parameter space. The transition due to general EFT kink
is suppressed  as
$\sim \exp(-\mu\lambda/\kappa)\sim \exp(-1/\alpha_s)$,
because $\kappa \sim \alpha_s^2 N_c$ (see appendix B). The
critical kinks that correspond to actual transitions between vacua
are the limiting cases of the families of these noncritical kinks.

\par
The transition to the black disk limit is of the inflationary
type: in the case of collisions of two small dipoles  the time
scale $T_I\sim \exp(1/\mu)/Q$ of the transition is significantly
smaller than the time needed for the formation of the perturbative
ladder $\propto 1/Qx^{1-\mu(Q^2)}$.

\par
The nonperturbative transition produces ladders that strongly
overlap in the impact parameter space.  Due to the exchange of the
constituents between the overlapping ladders the system of ladders
becomes the color network that resembles the lengthy but narrow
pencil.  The difference between the pencil and the jet is the
softer distribution of hadrons over transverse and longitudinal
momenta. However to study this color network, we need to be able
to describe the BDL transitions (the  kinks) directly in terms of the
QCD language, the goal that is yet not achieved.

\par
To summarize, we were able to show that QCD leads to solvable in
quasiclassical approximation effective effective field theory, and
this effective field theory  has a transition to the black disk
limit , that in the QCD language is a color network.

\par
The most challenging problems for the future work will be to
calculate both the kinks and the "phonons" of the EFT and to study
the  condensation of ladders directly in terms  of QCD
(quark,gluon) degrees of freedom and to understand the role of the
found QCD phenomena in two scale processes.

\acknowledgements{We thank M. Strikman  for useful discussions and
B. Svetitsky for the reading text and  comments.}
\appendix
\section{ Forces between the perturbative ladders.}

\par It has been known long ago \cite{Gribov1} that ladders are the dominant
degrees of freedom in the theoretical description of the high
energy processes. In the leading twist approximation this
assumption has been proved in pQCD. At the same time near the
black disk limit due to the existence of the triple ladder
coupling the multiladder configurations become important and may
overlap. We explain here that there are no long range forces
between the pQCD ladders and quarks and gluons are confined within
the ladders, till the latter start to overlap  inside the
space-time.

\par Indeed, consider the DGLAP ladder. The vector potential
created by this ladder can be calculated as
\beq
A^{a}_\mu(x)=\int d^4y D^{ab}_{\rm ret\,\,
\mu\nu}(x-y)J^b_\nu(y)
\label{A1}
\eeq
Here $D_{\rm ret}$ is the retarded gluon propagator
 in a light-cone gauge, while J is the matrix element of a current for gluon emission
by a ladder . The equation (\ref{A1}) can be rewritten in the
momentum space as
 \beq
 A^{a}_\mu(x)=\int d^4s \exp(ixs) D^{ab}_{\rm ret\,\,
 \mu,\nu}(s) J^b_{\nu}(s)
 \label{A2}
 \eeq
The retarded propagator in the light cone gauge has the form: \beq
D_{\rm ret}(s)=\frac{1}{s+i\epsilon
s_0}(g_{\mu\nu}-(s^\mu\eta^\nu+s^\nu\eta^\mu)/(s\eta)). \label{A3}
\eeq As usual  we take into account for the cancellation
 of the most singular term in the propagator that follows from  the Ward identities.
$J^b_{\nu}(s)$ is the current of a soft  gluon emission by the
ladder and $\eta$ is a light cone vector. \beq
J_{\mu}(s)=\sum^{i=N}_{i=1} q^{\mu}_i/(sq_i) \label {A4} \eeq
where the sum is taken  over all external lines of a cut ladder.
The contributions from the internal lines cancel at least in the
leading order \cite{Catani} . Consider integral over s, i.e. $\int
ds_-ds_+d^2s_t$.  Calculating the integral over the residues , we
see that the integral is controlled by $1/s^2$ pole in the
propagator. Potentially dangerous terms like $1/(q_i-s)^2$ are far
from the mass shell and have no singularity . Thus integral for
the vector potential is given by the pole $s^2=0$. The emitted
gluon is always on the mass shell and transverse. In other words
no long range forces exist within the DGLAP approximation.

\par The same arguments can be applied to BFKL ladder. The
integral over s is also determined by $1/s^2$ pole, and leads to
transverse gluons (ref. \cite{kovchegov}).

\par We conclude that the leading order ladders do not create perturbatively long-range fields in the leading twist approximation.
Generalization of this statement to the next-to-leading order
should be straightforward.

\section{Many "Pomeron" couplings in pQCD.}

The conventional strategy can be used to evaluate parameters
of effective Hamiltonian of EFT:  substitution of  variables in the path
integral from quarks and gluons to ladders. However such
calculation is too cumbersome at present. So we restrict ourselves
by  the qualitative evaluation of multi-ladder vertices  near the
unitarity limit in the lowest order of pQCD. The most difficult point
is to take into account for causality, i.e. location of singularities in the complex plane of energies in respect to the contour integration. It
has been shown by S.Mandelstam \cite{Mandelstam} and in the more
straightforward way by V.Gribov \cite{GribovCalc} that the
diagrams having no third spectral function $\rho(s,u)$  in Mandelstam
representation for the scattering amplitude give no contribution
into the leading power of energy. The transparent interpretation
of this result is that bare particle may experience only one
inelastic collision within the semiclassical approximation.

Thus lowest diagram for triple reggeon vertex is given by the
triangle gluon loop with 6 gluon lines  attached. So

\beq \kappa=G_{3P}\propto \frac{\alpha_s^2 N_c}{\lambda},
\label{B1} \eeq cf. recent discussion in ref. \cite{Bartels}. To
evaluate dependence on $N_c$ it is useful  to represent vectors in
the color space in terms of color spinors and then to find the
disentaglment of color contours.

The account of causality (i.e. of the fact that the  bare particle
may have only one inelastic collision and any number of elastic
ones) shows that  the lowest order diagram for the four ladder
vertex corresponds to the attachments of 4 ladders to the 2 gluon loops.

\beq
G_{4P}\propto \frac{\alpha_s ^4 N_c^2}{ \lambda^3}.
\label{B2}
\eeq

Similarly as the consequence of  the causality the $n$ ladder
vertex is given in the lowest order over $\alpha_s$ by the
attachment of $n$ ladders to the $n-2$ gluon loops. Then

\beq
G_{nP}\propto\frac{\alpha_s^{2n-4} N_c^{n-2}}{\lambda^{n-2}}.
\label{B3}
\eeq

The presence of the multi-ladder  couplings does not change the
behavior of the system near the extremum. This is because  their
relative contribution is characterized by the parameters:

$$G_{4P}\phi^4/G_{3P}\phi^3\approx \alpha_s N_c $$

Here $\phi$ is the field of the quasiparticle estimated in the WKB
approximation as $\approx \mu/\kappa$. Similarly one may estimate
the contribution of the $n\ge 3$ ladder vertices:

$$G_{nP}\phi^n/G_{3P}\phi^3\approx \alpha_s^{n-3} N_c^{n-3}$$

Thus relative contribution of higher vertexes is suppressed by
the powers of $\alpha_s$.

\end{document}